\documentclass[review, letterpaper, number, sort&compress]{elsarticle}
\usepackage[utf8]{inputenc}
\usepackage{amsmath}
\usepackage{amsfonts}
\usepackage{amssymb}
\usepackage{graphicx}
\usepackage{geometry}
\usepackage{natbib}
\usepackage[]{hyperref}

\journal{Nuclear Instruments \& Methods in Physics Research Section A}
\begin{document}

\begin{frontmatter}
\title{Proton Irradiation Damage and Annealing Effects in ON Semiconductor\textsuperscript{\textregistered} J-Series Silicon Photomultipliers}

\author[lanl]{K.D.~Bartlett\corref{cor1}}
\cortext[cor1]{Corresponding author}
\ead{kbartlett@lanl.gov}
\author[lanl]{D.D.S.~Coupland}
\author[lanl]{D.~Beckman}
\author[lanl]{K.E.~Mesick}

\address[lanl]{Los Alamos National Laboratory, Los Alamos, NM 87545 USA}

\begin{abstract}
Silicon photomultipliers (SiPMs) have become popular light conversion devices in recent years due to their low bias voltage and sensitivity to wavelengths emitted from common scintillating materials. These properties make them particularly attractive for resource-constrained missions such as space-based detector applications. However the space radiation environment is known to be particularly harsh on semiconductor devices, where high particle fluences can degrade performance over time. The radiation hardness of a particular SiPM, manufactured by ON Semiconductor\textsuperscript{\textregistered} (formally SensL), has yet to be studied with high energy protons, which are native to the space radiation environment. To study these effects we have irradiated groups of two SiPMs to four different fluences of 800 MeV protons delivered by the accelerator at the Los Alamos Neutron Science Center. Fluences of $1.68\times10^{9}$, $1.73\times10^{10}$, $6.91\times10^{10}$, and $1.73\times10^{11} \text{ protons}\text{ cm}^{-2}$, and their corresponding estimated doses of 0.15, 1.55, 6.19, and 15.5 kRad, were chosen based on estimates of the potential exposure a SiPM might receive during an interplanetary space mission lasting 10 years. We report the effects these doses have on dark current and the self-annealing time.
\end{abstract}


\end{frontmatter}

%\linenumbers
\section{Introduction}
Silicon photomultipiers (SiPMs) have become popular light conversion devices for nuclear particle instrumentation. Recently they have been used as direct replacements for more traditional technologies such as vacuum photomultipier tubes (PMTs). SiPMs have several advantages over legacy technologies which include: low bias voltage (on the order of tens of volts), compact size, low cost made possible by the modern manufacturing techniques of the semiconductor industry, reasonable temperature stability, comparable signal gains to that of PMTs, insensitivity to magnetic fields, and high photon detection efficiency over a range of wavelengths emitted from common scintillators. 

At a high level, SiPMs are silicon semiconductor devices that consist of a meshed network of Geiger mode avalanche photodiodes. They typically have active areas in the range of a few square millimeters and thus need to be tiled into larger arrays with customized front-end electronics before they can be effectively used as replacements to alternative technologies, like PMTs. A major disadvantage of SiPMs is their vulnerability to damage from prolonged radiation exposure, which initially leads to increased levels of dark current and corresponding dark count rates~\cite{GARUTTI201969}.

For space-based particle instrumentation, SiPMs are an attractive alternative to PMTs due to their low size, weight, and power (SWaP) requirements. There are several current and upcoming space missions that are making use of SiPM technology in their instrumentation packages. These include a Los Alamos National Laboratory (LANL) national security instrument scheduled to fly in 2020 at geosynchronous orbit~\cite{SENSER} and a prototype neutron and gamma-ray detection instrument called the Elpasolite Planetary Ice and Composition Spectrometer (EPICS)~\cite{EPICS}, Naval Research Laboratory instruments SIRI~\cite{SIRI,SIRIEarlyResults}, SIRI-2~\cite{SIRI-2} and future instrument concept Glowbug~\cite{Glowbug}, and the CsI instruments on NASA's AMEGO mission~\cite{AMEGO} and Burstcube mission~\cite{Burstcube}. 

In the space environment, instrumentation can receive a significant dose of ionizing radiation from proton fluxes in the form of galactic cosmic rays (GCRs), solar proton events, and trapped protons in earth's radiation belts. GCRs mainly consist of higher energy protons ranging from 1 MeV to 1 TeV with a peak centroid of 100s of MeV. Solar protons can carry energies greater than 100 MeV. Over the course of a 10 year interplanetary space mission, instrumentation can receive total protons fluences on the order of $10^{9}$ to $10^{10}$ protons cm$^{-2}$ from GCRs and solar proton events respectively. For earth orbiting missions these fluences can be 3 to 4 orders of magnitude larger, depending on the orbit, due to trapped protons in belts with energies between 10 MeV and 400 MeV, as estimated from the AP8 model~\cite{AP8Model}.

A major cause for concern for most of these mentioned missions is the performance degradation due to long term proton radiation exposure. Radiation effects due to hadrons, namely protons and neutrons, in SiPMs from a variety of manufacturers have been investigated~\cite{QIANG2013234,SANCHEZMAJOS2009506,HEERING2016111,MATSUMURA2009301,BOHN2009722,LI201663,MUSIENKO200987,6829721}, but have been found to vary by manufacturer, model, and dose particle species. An excellent overview of radiation damage mechanisms in SiPMs and current measurements, as recent as May, 2019, is given in~\cite{GARUTTI201969}. Proton irradiation and annealing effects in ON Semiconductor\textsuperscript{\textregistered} J-series SiPMs, which are planned to be or are currently being used in several of the previously mentioned space missions, have yet to be widely studied and reported in the literature. 

We report the first results of a proton irradiation study performed on J-series SiPMs manufactured by ON Semiconductor\textsuperscript{\textregistered} (formally SensL), which includes measurements of the dark current pre- and post-irradiation and a measure of the self-annealing time at room temperature. 

\section{Experimental Method}
We purchased eight $6 \text{ mm} \times 6 \text{ mm}$ active area J-series SiPMs, model number MicroFJ-SMTPA-60035-GEVB, from ON Semiconductor\textsuperscript{\textregistered} for the purpose of investigating their proton radiation hardness. The samples were taken for study to the Los Alamos Neutron Science Center Target 2 facility (Blueroom) on October 28th of 2018. The SiPMs were divided into four groups of two, and irradiated with $3.9\times10^{9}$, $4\times10^{10}$, $1.6\times10^{10}$, or $4\times10^{11}$ 800 MeV protons. The total number of protons in the beam were converted to fluences based on knowledge of the beam's cross-sectional profile. An effort was made during the experiment to measure the profile of the beam using a photographic plate; however, a single proton pulse from the accelerator saturated the plate and only yielded an estimate of the beam's overall diameter. Estimates of the fluence were thus made from calculations assuming a Gaussian beam profile with a 1 cm FWHM width, which was a choice informed from previous unpublished measurements. Additionally, the estimated dose was calculated from the fluence using a Geant4~\cite{AGOSTINELLI2003250,1610988,ALLISON2016186} simulation. The total number of protons, calculated fluences, and estimated dose for each of the samples are given in Table~\ref{tab:irradiation_level}.

\begin{table}[b]
	\centering
	\caption[Irradiation Level]{Summary of the total number of protons delivered per irradiation level, the calculated fluence incident on each sample, and the estimated dose per sample as determined from simulation.}
	\label{tab:irradiation_level}
	\begin{tabular}{ccccc}
	\hline
	Irradiation & Sample & Protons & Fluence on & Estimated \\
	Level & Label & Delivered & Sample [protons cm$^{-2}$] & Dose [kRad] \\ \hline
	1 & SiPM-0/SiPM-1 & $3.9\times10^{9}$ & $1.68\times10^{9}$ & $0.150$ \\
	2 & SiPM-2/SiPM-3 & $4\times10^{10}$ & $1.73\times10^{10}$ & $1.55$ \\
	3 & SiPM-4/SiPM-5 & $1.6\times10^{11}$ & $6.91\times10^{10}$ & $6.19$ \\
	4 & SiPM-6/SiPM-7 & $4\times10^{11}$ & $1.73\times10^{11}$ & $15.5$ \\ 
	\hline
	\end{tabular} 
\end{table}

\begin{figure}[t]
	\centering
	\includegraphics{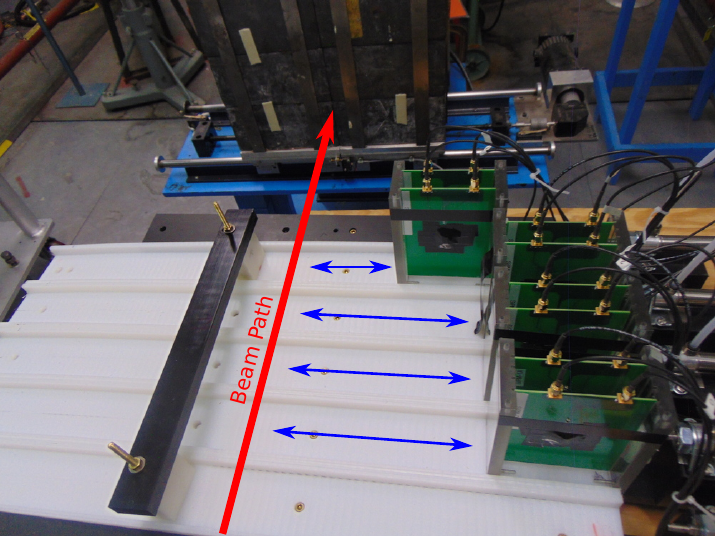}
	\caption{Pneumatic actuation device used to place the SiPM mounted PCB cards into the path of the beam (red arrow). Blue arrows represent the motion of travel of the card carriages.}
	\label{fig:SiPMpneumaticactuation}
\end{figure}

Each SiPM sample was solder mounted to a 4 inch by 6 inch multi-layer PCB card. These cards provided an easy pin-out configuration that allowed the SiPMs to be biased and read out in a compact form factor. Multiple layers of black electrical tape were applied to the surface, over the SiPMs, and backs of the PCB cards in an effort to shield out the ambient light. No light leaks were found in laboratory measurements prior to this study. Additionally, theses boards were used as support structures for the SiPMs as they were placed into the path of the proton beam. A pneumatic actuation system was used to move the SiPM mounted PCBs into and out of the beam's path, as shown in Figure \ref{fig:SiPMpneumaticactuation}. The actuation system was placed on top of a leveling table with global castor wheels, which gave us the ability to align the system with the beam. Alignment of the SiPMs was verified with a laser temporarily located in the path of the beam. Each card carriage was moved into and out of place to check that the beam was centered on each of the SiPMs. The castor wheels on the leveling table were then locked, affixing the system in place for the duration of the experiment. During the irradiation process the SiPMs were continuously biased at 27.5 V in order to replicate the operational conditions of a real space mission.

The irradiation process began with all of the card carriages placed into the path of the beam. An initial series of pulses with a number of protons equivalent to the first irradiation level given in Table~\ref{tab:irradiation_level} were delivered to all eight samples. Then the first card carriage was retracted, removing SiPM-0 and SiPM-1 from the path of the beam. An additional series of beam pulses, leading to a total equivalent exposure to the second irradiation level, was delivered to the remaining six SiPMs located in the beam's path. This process was then repeated for the remaining two levels of irradiation. Upon completion of the experiment the SiPMs were de-biased and removed from the experimental area, inside of the accelerator enclosure, and were placed into cold storage in an effort to reduce the amount of uncontrolled annealing until they could be moved back to our laboratory. A practice that was adopted based on previously published observations of annealing at cold temperatures~\cite{QIANG2013234}. A simple commercially available chest freezer kept the SiPMs at approximately $-23.0\pm1.0 ^{\circ}\text{C}$ during that initial cold storage period. 

Once the SiPMs returned to our laboratory, we began long term monitoring to characterize the annealing behavior of the samples at different temperatures. The even numbered SiPMs (SiPM-0, SiPM-2, SiPM-4, SiPM-6) were kept out at room temperature ($22.5\pm2.0^{\circ}\text{C}$) for the duration of the annealing study, while the odd numbered SiPMs (SiPM-1, SiPM-3, SiPM-5, SiPM-7) were placed into a Tenney Jr. thermal chamber at $-20.0\pm0.5^{\circ}\text{C}$ for the remainder of the study. Again the SiPMs were biased to 27.5 V for the duration of this monitoring period.

The main diagnostic method used in this study was measurements of the dark current versus applied bias voltage, also known as I-V curves. During these measurements dark currents were measured using a Keithley electrometer (model number 6514), while the bias voltage was provided by a Keithley three channel DC power supply (model number 2230G-30-1). Prior to irradiating the samples, a baseline I-V curve was measured for each SiPM in the laboratory at room temperature. Post-irradiation spot measurements, at a few select bias voltages, where made on October 28th (approximately 30 minutes after irradiation) at room temperature to get the immediate increase in dark current. Full I-V curve measurements were then made on November 5th, November 13th, November 19th, and November 26th, which approximately correspond to 188 hours, 380 hours, 524 hours, and 692 hours post-irradiation, respectively. During the full I-V curve measurements in November, the even numbered SiPMs were kept at room temperature while the odds were at $-20^{\circ}\text{C}$. 

On approximately December 3rd, an electrical failure in our biasing and monitoring system damaged the SiPMs and prematurely ended the study.

\section{Results \& Discussion}

\subsection{Dark Current}
In the time leading up to the irradiation study, the eight SiPMs were characterized in our laboratory for their baseline performance. Dark current for each of the eight samples was recorded from the maximum bias voltage of 27.5 V to 24.0 V in 1.0 V steps over the range of 27.5 V to 25.5 V, 0.5 V steps from 25.5 V to 25.0 V, and 0.1 V steps from 25.0 V to 24.0 V. A decreasing step size was used to accurately map the dark current about the breakdown voltage point, which was informed by the specification and performance documentation provided by the manufacturer~\cite{OnSemiSeriesJDataSheet}. I-V curves measured post-irradiation had a voltage step size of 0.1 V so that both the maximum and breakdown voltages could be accurately mapped.

Dark currents recorded from the initial characterization and post-irradiation measurements are tabulated in the third and fourth columns of Table~\ref{tab:postirrad_sipm_comparison}. These values are quoted at a bias voltage of 25.5 V and were chosen as a basis of comparison as it corresponded to values of dark current that were still within the measurement limitations of the Keithley electrometers, which had a maximum ceiling of 10 mA based on the range precision settings used, as measured across all of the SiPM samples post-irradiation. A conservative 5\% uncertainty was applied to the measured dark currents due to the fact that the values reported by the Keithley electrometer were unstable even in a slow 100 ms averaging mode.

\begin{table}[b]
	\centering
	\caption{Dark current before (Oct. 26) SiPM irradiation and immediately after (Oct. 28) irradiation at a bias voltage of 25.5 V.}
	\label{tab:postirrad_sipm_comparison}	
	\begin{tabular}{cccc}
	\hline
	& Estimated & \multicolumn{2}{c}{Dark Current {[}mA{]}} \\
	SiPM & Dose [kRad] & Pre-irradiation & Post-irradiation \\ \hline
	0 & 0.150 & 0.000162 & 0.114 \\
	1 & 0.150 & 0.000187 & 0.137 \\
	2 & 1.55 & 0.000138 & 0.347 \\
	3 & 1.55 & 0.000257 & 0.371 \\
	4 & 6.19 & 0.000196 & 0.861 \\
	5 & 6.19 & 0.000154 & 0.697 \\
	6 & 15.5 & 0.000237 & 2.470 \\
	7 & 15.5 &0.000195 & 1.800 \\ \hline
	\end{tabular}
\end{table}

\begin{figure}[t]
	\centering
	\includegraphics{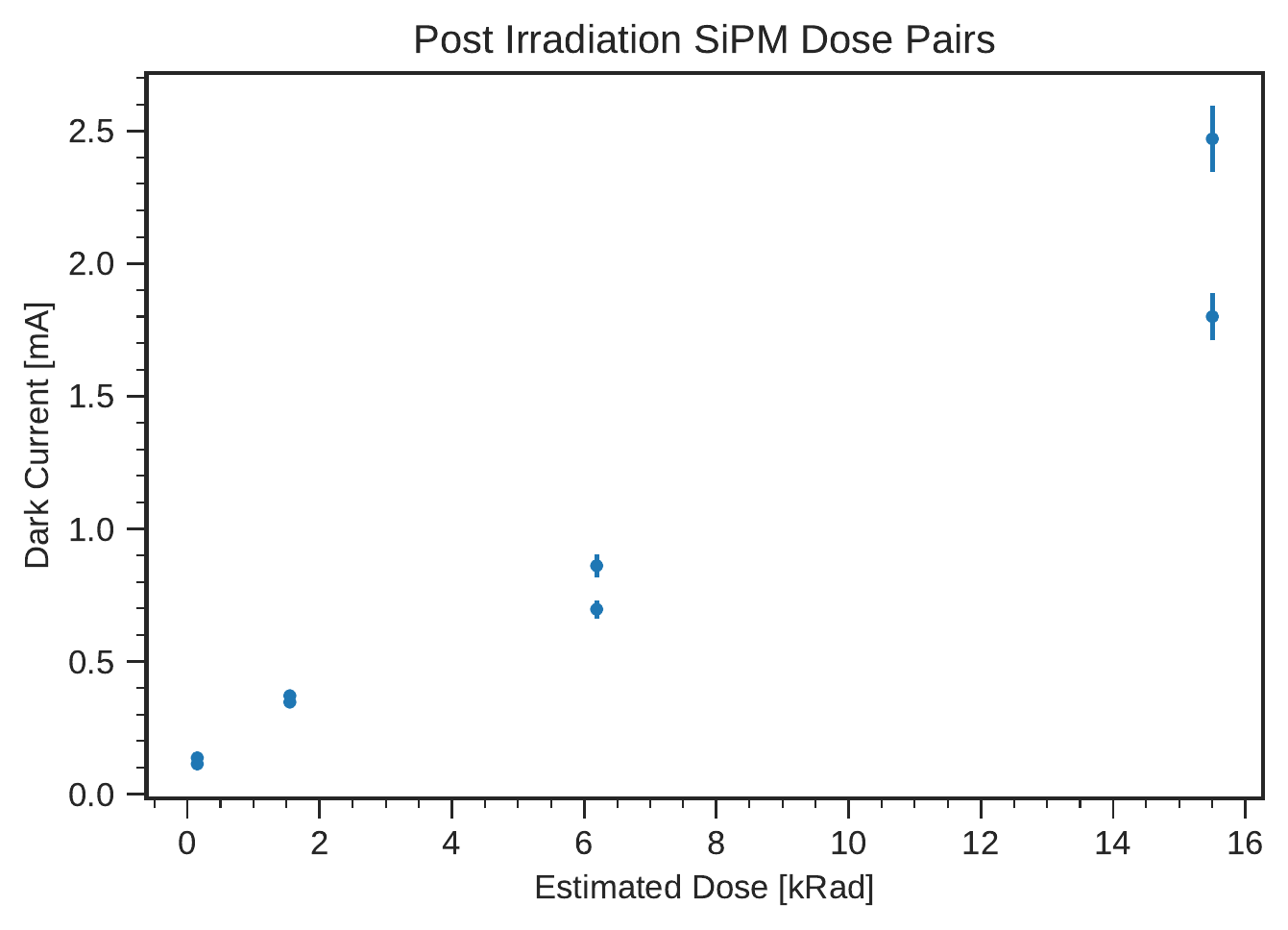}
	\caption{Comparison between dark currents measured, at 25.5 V bias voltage, directly after irradiation for pairs of SiPMs with common irradiation levels.}
	\label{fig:postirrad_sipm_comparison}
\end{figure}

Comparing the pre-irradiated and post-irradiated currents in Table~\ref{tab:postirrad_sipm_comparison}, we observed dark current increases by factors of approximately 718, 1980, 4460, and 9830 for doses of 0.15, 1.55, 6.91, and 15.5 kRad, respectively. These values are at least an order of magnitude larger than previously reported observations~\cite{GARUTTI201969} for SiPMs from a variety of manufacturers exposed to different hadron species. In particular, one previous study observed only a factor of 23 increase for a neutron fluence corresponding to approximately to our lowest proton fluence~\cite{6551130}.

Pairs of SiPMs irradiated to a given dose level agree well for the lowest two levels, but diverge for the last two levels, as seen in Figure~\ref{fig:postirrad_sipm_comparison}. This difference could reflect variability in the response of the different SiPMs to radiation damage or it could be the result of some other effect. It is important to note that the pre-irradiated SiPMs have approximately a 20\% variation in their measured dark currents, which could be contributing to the observed response post-irradiation. Further study with a larger sample size could determine if this observation is a statistical fluctuation or a true response in these particular make of SiPMs. 

\begin{figure}[t]
	\centering
	\includegraphics{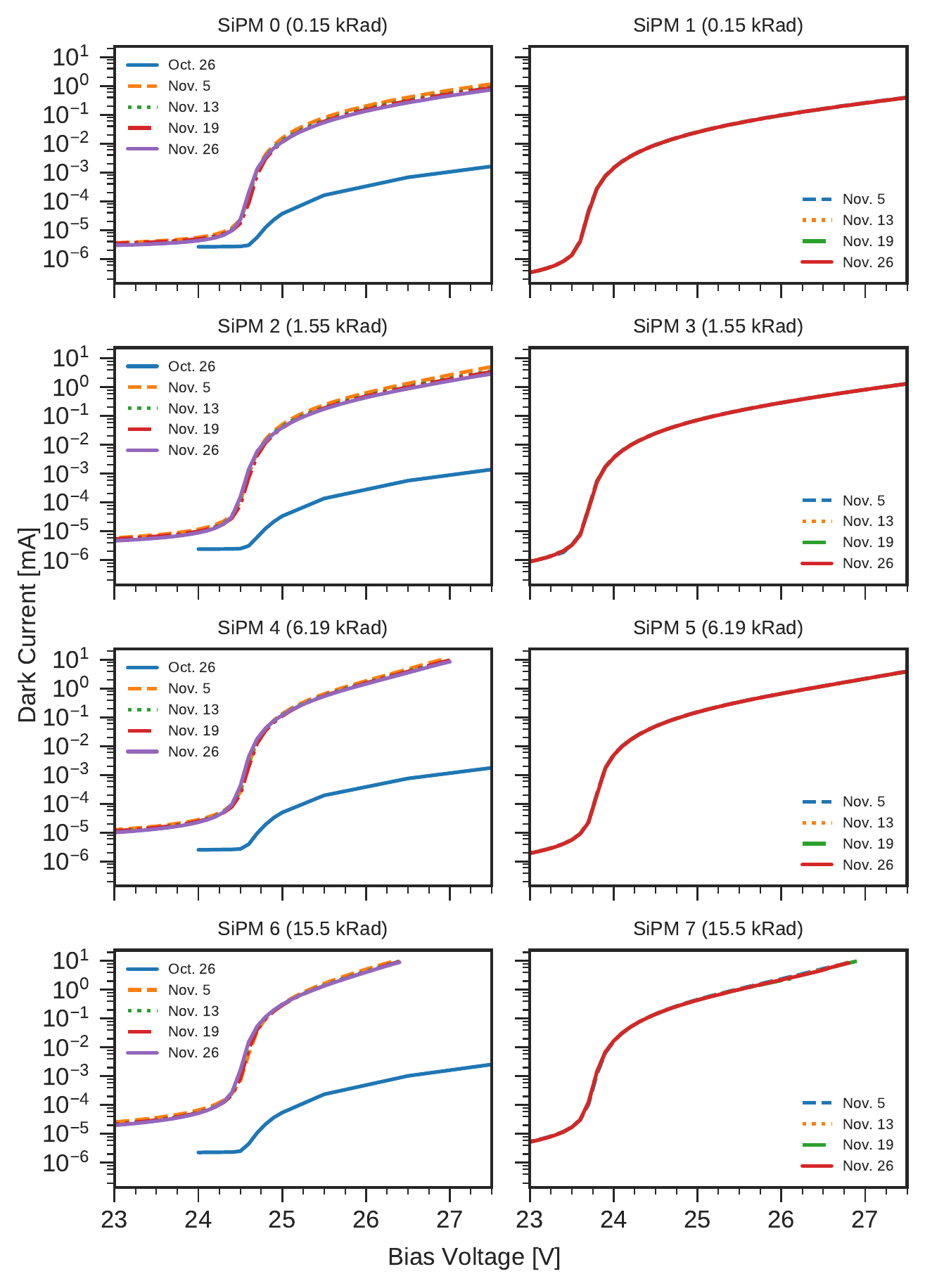}
	\caption{Pre-irradiation (Oct. 26) I-V curves for the even numbered SiPMs and post-irradiation (Nov. 5 through Nov. 26) I-V curves for all the SiPMs. Curves for the even numbered SiPMs were measured at room temperature, while the odd ones were measured at $-20\pm0.5^{\circ}\text{C}$.}
	\label{fig:IVcurves}
\end{figure}

Pre-irradiation I-V curves for the even numbered SiPMs and post-irradiation curves for all of the SiPMs are presented in Figure~\ref{fig:IVcurves}. Focusing on the even numbered SiPMs, it appears their breakdown voltages pre- and post-irradiation are consistent within a few tenths of a volt. Any difference at first order can be easily described by the temperature dependence of the the breakdown voltage, which is approximately 21.5 mV/$^{\circ}$C~\cite{OnSemiSeriesJDataSheet}. The temperature varied a few degrees centigrade between the pre- and post-irradiation I-V curve measurement, which was due to the natural variation in the lab's ambient atmosphere.

The general structure between the pre- and post-irradiation I-V curves shows a change in the slope of the breakdown voltage turnover point, which has become more gradual as the background dark current has increased across the entire bias voltage range. 

Small differences, on a log scale, seen in the dark current for the even numbered SiPMs at room temperature in Figure~\ref{fig:IVcurves} is an indication of annealing or recovery to a new baseline dark current.

\subsection{Annealing}
As previously mentioned, I-V curves were measured throughout the month of November in an attempt to characterize the amount of annealing occurring both at room temperature and at $-20^{\circ}\text{C}$. The measured dark currents recorded at $-20^{\circ}\text{C}$ were consistent with a constant value which is an indication that an undetectable amount of annealing had occurred. However, a month was a long enough duration of time to observe and characterize annealing at room temperature. 

A simple exponential decay function with three free parameters, in the form of
\begin{equation}\label{eq:annealtimefit}
f(t; A, \tau, C) = Ae^{-t/\tau}+C,
\end{equation}
\noindent was chosen to model the exponential decay in the dark current during that month period based on previously published results~\cite{QIANG2013234}. The three free parameters in the model are the amplitude ($A$), decay constant ($\tau$), and offset ($C$), which are in units of milliamperes, hours, and milliamperes, respectively. These parameters can be interpreted as: $A$ the amount of possible dark current change with annealing, $\tau$ the characteristic annealing time, and $C$ the new baseline dark current after fully annealing at a given temperature. The LMFIT module~\cite{python_lmfit} for Python3 was used to perform the fits.

\begin{table}[b]
	\centering
	\caption{Exponential decay model parameters for room temperature annealing extracted from fits to dark currents measured at 25.5 V bias voltage.}
	\label{tab:annealingfitparameters}
	\begin{tabular}{ccccccc}
	\hline
	 & Estimated & Amplitude & Decay & Constant & $\chi^{2}$ & Fit \\
	SiPM & Dose [kRad] & (A) [mA] & Constant ($\tau$) [hrs] & (C) [mA] & Fit Stat. & Prob.\\ \hline
	0 & 0.150 & $0.063 \pm 0.002$ & $237.181 \pm 17.635$ & $0.051 \pm 0.001$ & $2.1\times10^{-5}$ & 0.999 \\
	2 & 1.55 & $0.176 \pm 0.007$ & $223.217 \pm 18.800$ & $0.168 \pm 0.003$ & $7.1\times10^{-5}$ & 0.999\\
	4 & 6.19 & $0.329 \pm 0.012$ & $188.183 \pm 13.910$ & $0.529 \pm 0.005$ & $1.1\times10^{-4}$ & 0.999 \\
	6 & 15.5 & $1.118 \pm 0.002$ & $142.509 \pm 0.386$ & $1.352 \pm 0.001$ & $6.2\times10^{-7}$ & 0.999\\ \hline
	\end{tabular}
\end{table}

\begin{figure}[t]
	\centering
	\includegraphics{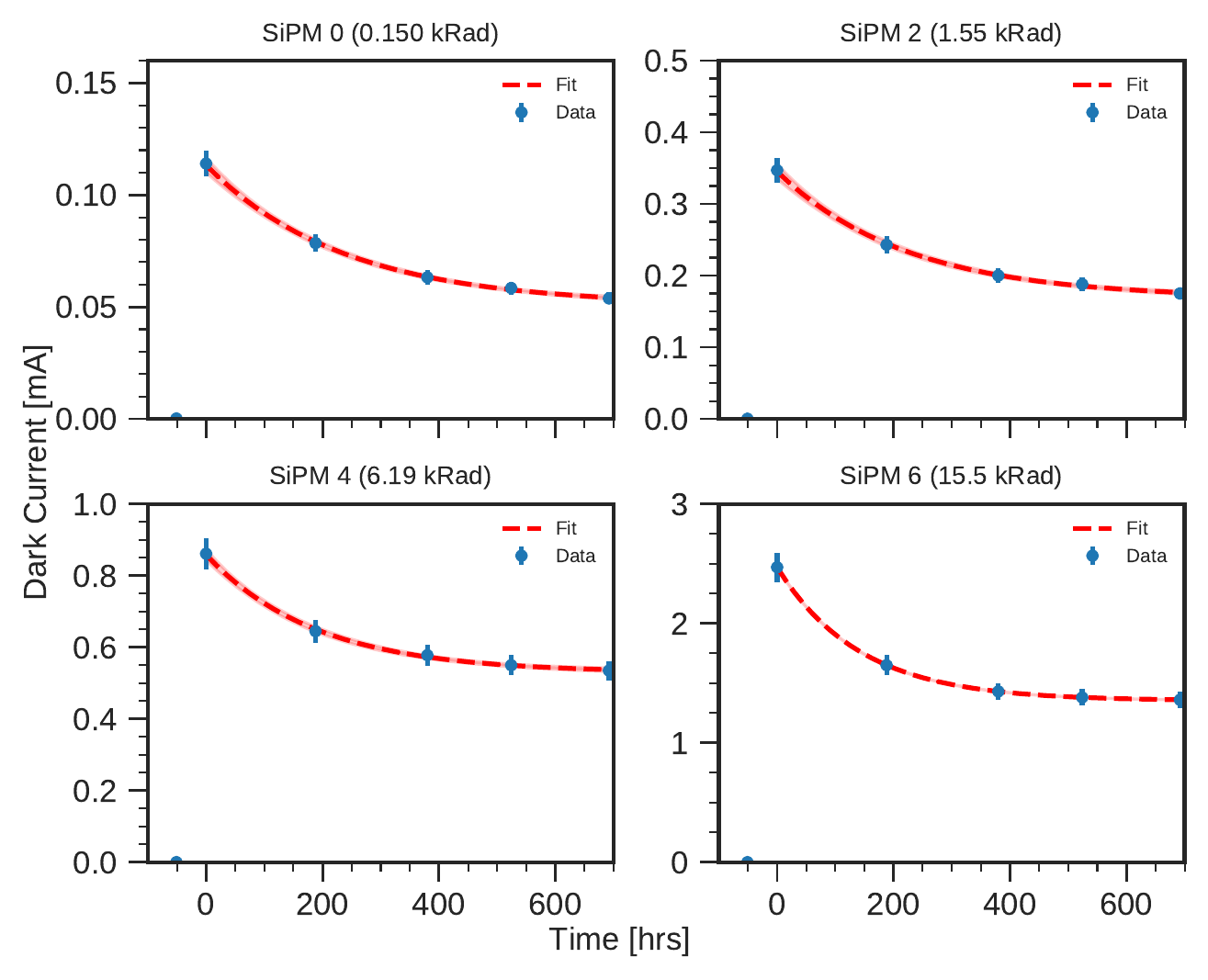}
	\caption{Dark currents (blue) measured at 25.5 V bias voltage from the SiPMs kept at room temperature, superimposed with an exponential decay fit (red) used to extract annealing times.}
	\label{fig:annealingfits}
\end{figure}

Results from the fits and the fit statistics are given in Table~\ref{tab:annealingfitparameters}. Additionally, a superposition between the fits and data are given in Figure~\ref{fig:annealingfits} for ease of comparison. Overall, the fits do an excellent job of modeling the decay in the data based on the reported $\chi^{2}$ fit statistic and probability from the fit routine. In fact the fit statistics report an overfitted result, indicating the conservative uncertainty assigned to the data is an overestimate and could be reduced. 

The lowest two irradiation levels had annealing times, extracted from the fits, of approximately 10 days (240 hrs) and these times decreased in a general linear trend with dose to approximately 6 days (144 hrs) for the highest irradiation level. This result is consistent with previously published results where they observed a self-annealing time of approximately 10 days in a similar SiPM from the same manufacturer, which received a neutron fluence of $3.7\times10^{9}\text{ neutrons}\text{ cm}^{-2}$~\cite{QIANG2013234}. 

At the end of the month the even numbered SiPMs, numbers 0, 2, 4, 6, saw an approximate 50\% decrease to their currents to 0.0538 mA, 0.175 mA, 0.535 mA, and 1.36 mA, respectively. These values are close to the theoretical post annealing baseline dark currents determined by the fits (offset C), which are 0.051 mA, 0.168 mA, 0.529 mA, and 1.352 mA for SiPMs 0, 2, 4, and 6, respectively. Although a 50\% reduction in the dark current post-irradiation from annealing at room temperature was observed, the dark currents remain orders of magnitude above their pre-irradiated levels. This size of recovery in dark current is in agreement with the previously published and mentioned result~\cite{QIANG2013234}. It is important to note that other investigations seem to indicate that increased temperatures could aid in this annealing process~\cite{GARUTTI201969} and should also be investigated in the future for J-series SiPMs.

\section{Summary}
We investigated the effects of proton radiation damage and annealing in ON Semiconductor\textsuperscript{\textregistered} J-series SiPMs for the first time. Eight samples were irradiated by 800 MeV protons at the Los Alamos Neutron Science Center. Groups of two SiPMs were irradiated to four increasing fluence levels, which were approximately $1.68\times10^{9}$, $1.73\times10^{10}$, $6.91\times10^{10}$, and $1.73\times10^{11} \text{ protons}\text{ cm}^{-2}$. These fluences correspond to doses of 0.150 kRad, 1.55 kRad, 6.19 kRad, and 15.5 kRad, respectively. The lowest level investigated corresponds to the estimated fluence a SiPM could receive from GCR protons over the course of a 10 year interplanetary space mission, where as the higher levels correspond to potential fluences received from solar protons over the course of the same type of mission.

Effects from the irradiation process were investigated using measurements of the SiPMs dark current versus bias voltage. Immediately post-irradiation the SiPMs exhibited a dramatic increase in the amount of dark current. At the lowest level of dose we observed an approximate factor of 715 increase from the pre-irradiated baseline and approximately a factor of 10,000 increase for the highest dose. Comparing to previous published results our observations are orders of magnitude higher than the former, which have primarily used neutrons to study displacement damage effects at much lower energies. 

Post-irradiation we attempted to measure the annealing time required to partially heal the high dark current behavior at multiple temperatures. A failure in our experimental setup caused the annealing investigation to be cut short at only one month's time. Only the SiPMs held at room temperature showed a large enough decrease in the amount of dark current that an annealing time could be extracted using an exponential decay model, whereas the SiPMs at $-20^{\circ}\text{C}$ did not show change in the dark current over the one month monitoring period. Approximately 50\% of the original post-irradiation dark current was recovered during this annealing period for the even numbered SiPMs. An inverse relationship was observed between these annealing times and received dose that ranged from approximately 10 to 6 days. Additionally, our extracted annealing times agree well with previously measured results that used SiPMs from the same manufacturer.

SiPMs are susceptible to radiation damage over the range of fluences and corresponding doses investigated here, as indicated by their increased levels of dark current. A measurable amount of the damage can be recovered via annealing, which is a temperature dependent process. Further study with a larger sample size is required to quantify how much of the damage can be recovered in J-series SiPMs at temperatures other than room temperature.

\section*{Acknowledgments}
The authors would like to thank Michael Mocko, Matt Devlin, Steve Wender, Erik Krause, Jeff George, and the remaining members of the staff of the LANSCE Blueroom for their helpful support with this study. The work presented in this publication was supported by the Los Alamos National Laboratory: Laboratory Directed Research and Development program under project number 20170438ER. 

\bibliography{references}

\end{document}